# Learning From Others, Together: Brokerage, Closure and Team Performance

JOSE URIBE AND DAN WANG, Columbia Business School

## 1. INTRODUCTION

Scholarship on teams has focused on the relationship between a team's performance, however defined, and the network structure among team members. For example, Uzzi and Spiro (2005) find that the creative performance of Broadway musical teams depends heavily on the internal cohesion of team members and their past collaborative experience with individuals outside their immediate teams. In other words, team members' internal cohesion and external ties are crucial to the team's success. How, then, do they interact to produce positive performance outcomes? In our work, we separate the proximal causes of tie formation from the proximal determinants of outcomes to determine the mechanism behind this interaction. To examine this puzzle, we examine the performance of national soccer squads over time as a function of changing levels and configurations of brokerage and closure ties formed by players working for professional soccer clubs.

National soccer players' membership networks are formed in two different types of teams, referred throughout this article as "clubs" and "squads". Clubs are professional teams that hire players domestically and internationally to compete in domestic leagues such as the *MLS* in the U.S., the *English Premier League* or Germany's *Bundesliga* and in continental tournaments such as the *UEFA Champions League*. Each year, clubs must release players selected by their home country's national squad (which consist exclusively of citizens) to compete against other national squads in tournaments such as the *AFC Asian Cup*, the *Africa Cup of Nations*, and the *FIFA World Cup*. When players reassemble into their national squads, club-based membership networks create rich combinations of brokerage and closure.

## 2. BROKERAGE, CLOSURE AND TEAM ARCHITECTURE

Although several past work addressed the contingent effects of brokerage and closure (Burt 1997; Wang 2014; Xiao and Tsui 2007), few have identified specific structures that can best combine the benefits of both (Reagans and Zuckerman 2001; Uzzi and Spiro 2005). We uncover two basic team structures that can combine internal ties and external ties. Both the general advantages of brokerage and closure's contributions to performance when brokerage is high accrue through Simmelian bridges (Tortoriello and Krackhardt 2010), a specific structure that combines brokerage and closure within the same sub-teams. Theory on transactive memory systems (Argote and Ingram 2000; Ren and Argote 2011) and on social capital formation (Burt 2000; Burt 2005; Burt 2010) underpins our reasoning for why colocation of brokerage and closure structures in sub-teams enhances team performance. These findings reveal the contingent nature of network ties in teams and points to a strategic role for network architecture in competitive environments. In addition, we provide the first empirical test of the importance of Simmelian bridges for team outcomes.

### 2.1 Brokerage

Research shows employee mobility to be an important factor in the transfer of tacit knowledge between organizations (Corredoira and Rosenkopf 2010; Song, Almeida and Wu 2003; Wezel, Cattani and Pennings 2006). We refer to *brokerage* as the presence of collaborative relationships team members have with members of other teams – i.e., ties that span national squad boundaries. Mobile employees accumulate both origin and destination-specific human capital through learning by doing at both loci (Carnahan and Somaya 2013). In essence, these external ties serve as pathways to knowledge resources in other teams. Several studies suggest that teams can perform more effectively if they obtain and use external knowledge, in the form of task-related information, know-how, and feedback from sources outside the teams (Haas 2010).  This process operates at a global scale in world soccer; players in highly internationalized clubs transfer valuable skills and knowledge to help their national squads' performance.

## 2.2   Closure

*Closure* – or the presence of dense collaborative interrelationships between team members formed outside their focal team – is typically associated with improved collective information processing and better implementation and coordination of tacit information, both critical drivers of team performance (Williams Woolley 2011).  These within-team ties typically enable members to spend more time interacting and learning from shared experiences.  Shared experiences lead groups to code, store, and retrieve information together, and enable the development of shared language and symbols, a precursor for effective communication (Cohen and Levinthal 1990).  Members sharing multiple teams can build a reservoir of tacit knowledge and relationship-specific heuristics crucial for processing non-codified information (Hansen 1999; Williams Woolley 2011).  High closure also facilitates collective information processing (Williams Woolley 2011) and transactive memory (e.g. Berman, Down and Hill 2002), which expands both the range of data attended to and the speed of processing (Uzzi 1997)

## 2.3   Team Architecture

Brokerage is thought to benefit team performance most when embedded in a dense social structure that facilitates the formation of common knowledge and shared meanings, reduces frictions, and promotes coordinated actions that are necessary to integrate and take advantage of diverse sources of knowledge (Bresman 2010; Burt 2005; Tortoriello and Krackhardt 2010).  Teams that combine diverse external ties – i.e., high brokerage – with dense internal ties – i.e., high closure – should perform best.  However, there is little guidance as to how much brokerage is required for the benefits of closure to take effect.  Moreover, scholarship is silent on the implications for teams that exhibit variation in the ways they organize internal and external ties.

We argue that the underlying characteristic determining the configuration of ties is the extent to which cohesive subgroups 'bridge together.'  Members sharing bridging ties with their own teammates form *Simmelian bridges.* Simmelian bridges were originally proposed by Tortoriello and Krackhardt  (2010) and defined as ties spanning internal boundaries in a formal organization that are surrounded by common third-party ties.  Simmelian bridges  combine the known advantages of brokerage (e.g., access to cutting-edge competitors and training) with the advantages of closure (e.g., stability, transactive memory, cohesion) (Tortoriello and Krackhardt 2010).  Conversely, structures that separate internal and external ties among different subgroups create an additional synapse where learning and skills acquired externally can be lost in translation.  In addition, those brokering alone, or *Solo bridges*, may face social friction and resistance from high-closure sub-teams, hampering the team's integration.

## 3.   METHODOLOGY

To construct the measures, we used player rosters for every national squad for each year between 1990 and 2010[1].  The final sample consisted of 16,092 unique players comprising 186 national squads over the period from 1990 to 2010.  The full set of players had affiliations to 4,037 clubs located in 224 countries. The dependent variable for team performance was the natural logarithm of squads' Elo rating on the following calendar year (i.e. 2001 team performance modeled with 2000 network measures). National squads' annual Elo ratings were constructed with data from http://www.eloratings.net/.

---

[1] We are immensely grateful to our late colleague Casey Ichniowski for generously sharing the database with us.

In the world soccer context, closure is the number of club-based ties between squadmates divided by the total number of possible in-group ties. Brokerage is the number of club-based ties with foreign nationals per player in a squad. We decompose brokerage into two mutually exclusive and collectively exhaustive types. Brokerage involving more than one focal squad player constitutes *Simmelian bridges*. Brokerage involving a single focal squad player is captured with the variable *Solo bridges*. Both *Solo bridges* and *Simmelian bridges* describe the manner in which team members learn and transfer skills and routines from abroad to their national squads.

We used linear regressions to model the impact of closure and brokerage on performance, measured as the squad's yearly average Elo rating during the 1990-2010 period. We controlled for the squad's past performance, domestic league strength, the percent of players in a squad affiliated with the top 30 European clubs in the previous season and clustered standard errors at the level of each squad to account for the persistence of performance within a team.

## 3.1   Results

We found that closure enhances performances at relatively high levels of brokerage. The main effect of closure throughout the models is negative, and closure reduces performance until brokerage reaches relatively high levels (about the 75th percentile in our sample). This effect persists after including controls for differences in club quality, the strength of the domestic league, and any unobservable path-dependent differences, captured by lagged performance.

After decomposing brokerage into Simmelian bridges and Solo bridges, we found that Simmelian bridges are a superior way to organize external ties. The only type of brokerage that contributes to performance in our models are Simmelian bridges. Moreover, the coefficient on the interaction between Simmelian bridges and closure is positive and significant, while the interaction between Solo bridges and closure, although positive, is not statistically significant. Closure's contribution to performance requires Simmelian bridges, which constitute a superior way of aligning brokerage and closure in teams, at least in a context of global competition and resource inequality.

## 4.   DISCUSSION AND CONCLUSION

We used performance and network data for national soccer teams from 1990-2010 to test how different levels of brokerage and closure help identify the best performers beyond other competitive advantages, such as having high quality players and an attractive domestic league. Our empirical setting is unique in that literally all sampled individuals lie at the intersection of the dual network of professional clubs and national squads. Among knowledge workers in the United States and Europe, between 65 to 95 percent belong to more than one project team at a time (Boyer O'leary, Mortensen and Williams Woolley 2011). Similarly, employee mobility has become the norm in knowledge work (Carnahan and Somaya 2013), and professional soccer players typically transfer every few years between clubs. The empirical strategy separates the origination of network ties from the locus where performance occurs, bypassing some of the perennial endogeneity and statistical identification concerns inherent to research on social networks. Individual memberships in professional soccer clubs and national soccer squads connect contexts under separate governance structures, enabling a systematic investigation on how collaborative networks may influence team performance.

Our findings suggests that when competition is global and the distribution of skills and resources unequal across teams, the performance returns to closure are contingent on both the amount of brokerage and the architecture of sub-teams. This study suggests that star power is not sufficient for team success. Rather, team outcomes are improved and sustained by combining rich collaboration between members with strategic contact with rivals through Simmelian bridges.